\documentclass{article}
\usepackage{graphicx}
\usepackage{amsmath,amssymb}
\usepackage{graphics,color,array,calc,rotating,epsfig,psfrag}
\usepackage{hyperref}
\numberwithin{equation}{section}
\usepackage{cite}
\usepackage{bm}
\usepackage{dcolumn}
\usepackage{float}
\usepackage{subfigure}
\usepackage{tikz-cd}
\usepackage{caption}
\usepackage[utf8]{inputenc}
\usepackage{tikz}
\usepackage{amsfonts}
\usepackage[mathscr]{eucal}
\usepackage{helvet}
\def\be{\begin{equation}} \def\ee{\end{equation}}
\def\bea{\begin{eqnarray}} \def\eea{\end{eqnarray}}

\newcommand{\reef}[1]{(\ref{#1})}

\newcommand{\nn}{\nonumber}
\newcommand{\RN}[1]{%
	\textup{\uppercase\expandafter{\romannumeral#1}}%
}
\begin{document}
	\baselineskip 18pt%
	\begin{titlepage}
		\vspace*{1mm}%
		\hfill%
		\vspace*{15mm}%
		\hfill
		\vbox{
			\halign{#\hfil         \cr
			} 
		}  
		\vspace*{20mm}
		\begin{center}
			{\Large {\bf \boldmath Marginal  $T \bar{T}$-like Deformation and ModMax Theories in Two Dimensions}}
		\end{center}
		\vspace*{5mm}
		\begin{center}
			{H. Babaei-Aghbolagh$^{\dagger}$, Komeil Babaei Velni$^{*}$,
				Davood Mahdavian Yekta$^{\ddagger}$ and H. Mohammadzadeh$^{\dagger}$
			}\\
			\vspace*{0.2cm}
			{\it
				$^{\dagger}$Department of Physics, University of Mohaghegh Ardabili,
				P.O. Box 179, Ardabil, Iran\\
				$^{*}$Department of Physics, University of Guilan, P.O. Box 41335-1914, Rasht, Iran\\
				$^{\ddagger}$Department of Physics, Hakim Sabzevari University, P.O. Box 397, Sabzevar, Iran\\
			}
			
			\vspace*{0.5cm}
			{E-mails: {\tt h.babaei@uma.ac.ir,  babaeivelni@guilan.ac.ir, d.mahdavian@hsu.ac.ir, mohammadzadeh@uma.ac.ir
			}}
			\vspace{1cm}
		\end{center}
\begin{abstract}
  Recently, the ModMax theory has been proposed as a unique conformal non-linear extension of electrodynamic theories. We have shown in \cite{Babaei-Aghbolagh:2022MoxMax} that this modification can be reproduced by using a marginal  $T \bar{T}$-like deformation from pure Maxwell theory. Further, it was shown that this deformation is solvable by applying a perturbative approach. In this paper, we will investigate a similar marginal $T \bar{T}$-like deformation for a general two-dimensional scalar field theory. It is shown that employing an irrelevant $T \bar{T}$ operator on this marginal scalar theory will produce a generalized Nambu-Goto action of this scalar theory which is a Born-Infeld-like action in two dimensions. Using a similar prescription for a two-dimensional theory with multiple scalar fields, we show that the marginal $T \bar{T}$-like deformation yields a ModMax-like Lagrangian and then the irrelevant operator produces a generalized scalar ModMax action.

\end{abstract}
\end{titlepage}
	\section{Introduction}\label{1}
	Recently, a precisely solvable deformation of two-dimensional (2D) quantum field theory (QFT) by an irrelevant operator, known as $T \bar{T}$-deformation, has been proposed by Zamolodchikov  and Smirnov \cite{Zamolodchikov:2004ce,Smirnov:2016lqw}. The $T \bar{T}$ deformation is an irrelevant operator that deforms  the free  theory with respect to a dimensionful parameter $\lambda$. This deformation is an explicit function of the corresponding energy-momentum tensor which is a well solvable theory in two dimensions and is defined as follows
	\begin{equation}
	\label{irr}
	{\cal L}\,(\lambda) ={\cal L}_{free} +\frac{1}{8} \int O_{T^2}^{\lambda} \,d\lambda\,,
	\end{equation}
	where, $O_{T^2}^{\lambda}$   is the standard irrelevant  $T \bar{T}$ operator in 2D field theory denoted by
	\begin{eqnarray}
	\label{olamd}
	O_{T^2}^{\lambda}= T_{\mu\nu}T^{\mu\nu}- {T_{\mu}}^{\mu} {T_{\nu}}^{\nu}\,.
	\end{eqnarray}
	 $T_{\mu\nu}$ is the energy-momentum tensor of the free theory.
	  In other words, the irrelevant perturbation in 2D field theory is addressed by the following flow equation \cite{Cavaglia:2016oda,Guica:2017lia}
		\begin{equation}
	\frac{\partial {\cal L}(\lambda)}{\partial \lambda}=\frac18\,  O_{T^2}^{\lambda}.
	\end{equation}
	Also, it has been shown that this kind of deformation has  some links to non-linear electrodynamics theory \cite{Conti:2018jho, Ferko:2019oyv, Babaei-Aghbolagh:2020kjg}, correlation functions\cite{Song He:2004oyv,Song He:2011oyv,Miao He:2109oyv,Song He:2202oyv}, dynamical coordinate transformations \cite{David,Gross:2019uxi,Song he:2104uxi} and AdS/CFT correspondence\cite{McGough:2016lol,Giveon:2017myj,Giribet:2017imm,Kraus:2018xrn,Donnelly:2018bef,Allameh:2021moy}.  
 Though this theory is solvable and has a finite spectrum in $D=2$, the deformed theory does not respect the conformal symmetry. Thus, it would be of interest to investigate another type of $T \bar{T}$-deformation that respects this symmetry.
	
    In $D=4$, a marginal modification of the Maxwell's theory with respect to the $SO(2)$ duality and conformal symmetry has been proposed in Ref. \cite{Bandos:2020jsw}. This electromagnetic modification is known as the ModMax theory. The generalization of the ModMax theory in the context of conformal invariant $p$-form gauge theories was shown in Ref. \cite{Bandos:2020hgy}. An alternative form of the ModMax Lagrangian in the presence of electrically and magnetically charged sources, such as axion-dilaton-like auxiliary scalar fields, has been studied in Ref. \cite{Sorokin:2206c}. We have shown in \cite{Babaei-Aghbolagh:2022MoxMax} that the ModMax theory is a marginal $T \bar{T}$-deformation of the Maxwell theory. In fact, the resultant ModMax action belongs to the flow of some typical marginal operator. Recently in Ref. \cite{Rodriguez:2021tcz}, it has been shown that when the conformal algebra in $D=2$ dimensions is mapped to its ultra/non-relativistic versions, the marginal operator can be appeared and thus, it can also be used to deform relativistic theories into ultra-relativistic ones. Also, it has been proposed in Ref. \cite{Bagchi:2022tcz} that there is a flow between 2D CFTs and Conformal Carrollian CFTs (or BMSFTs) in the same dimension via an infinite boost transformation, which is equivalent to deforming the theory with a marginal operator.
	
As a main purpose of our study in this paper, we investigate another type of $T \bar{T}$-like deformation, the so-called marginal $T \bar{T}$-deformation in 2D spacetime. This deformation includes a marginal operator which deforms the Lagrangian of the free scalar theory with respect to the dimensionless parameter $\gamma$. Since the usual $T \bar{T}$-like deformation by an irrelevant operator on 2D field theory yields the gauge-fixed Nambu-Goto Lagrangian \cite{Cavaglia:2016oda}, we will show that applying this kind of operator on the marginal scalar (MS) theory, one gets to a generalized 2D Nambu-Goto (NG) action similar to the generalized Born-Infeld (BI) action in 4D electrodynamics \cite{Babaei-Aghbolagh:2022MoxMax,Bandos:2020hgy}.

Our next motivation in this paper is to investigate the deformation of interacting classical action of a multi-scalar theory by employing a marginal operator. We show that the emergent deformed scalar field theory is very similar to the ModMax Lagrangian, therefore we named it the scalar ModMax (SMM) theory. It is of interest to point out in this respect that one can obtain the generalized scalar ModMax (GSMM) Lagrangian by using an irrelevant $T \bar{T}$ operator from a simple integration technique, which is a BI-like action of scalar fields in $D=2$.

This paper is organized as follows: In section \ref{2}, we first review the marginal $T \bar{T}$-like deformation of non-linear ModMax theory in $D=4$ dimensions. Then, we will propose a marginal operator for a massless scalar field theory in $D=2$ dimensions. Further, the non-linear extension of the emergent MS theory is investigated using an irrelevant  $T \bar{T}$ operator which is a generalized marginal scalar (GMS) theory.
In section \ref{3}, by iterating a similar prescription of the perturbation approach with a marginal $T \bar{T}$ operator on a multi-scalar theory, we obtain a non-trivial ModMax Lagrangian for a massless scalar fields theory in $D=2$. Then, by applying an irrelevant $T \bar{T}$ operator to this SMM theory, we obtain the generalized NG action of SMM theory. Section \ref{4} is devoted to giving a brief summary of results and outlook.
	\section{ Marginal  $T \bar{T}$-like deformation}\label{2}
The ModMax theory is a non-linear electrodynamic theory which is an extension of the Maxwell theory with respect to a dimensionless coupling constant $\gamma$. Under this modification, which is called the marginal deformation due to the dimensionless parameter of the deformation, the $SO(2)$ duality and the conformal symmetry are preserved \cite{Bandos:2020jsw,Bandos:2020hgy}.
	
	However, we have shown in Ref. \cite{Babaei-Aghbolagh:2022MoxMax} that there exists a marginal $T \bar{T}$ operator in the form of
	\begin{eqnarray}\label{Ogama}
	O^{\gamma }_{T^2}=\sqrt{T_{\mu\nu}T^{\mu\nu}- \frac{1}{4} {T_{\mu}}^{\mu} {T_{\nu}}^{\nu}}\,,
	\end{eqnarray}
	which reproduces the ModMax theory from the Maxwell electrodynamics by using the perturbative approach. In this section, we will review the marginal deformation by operator $O^{\gamma }_{T^2}$ in four dimensions and then find a 2D operator in analogous to the 4D case in Eq. \eqref{Ogama}. MS theory is the modified Lagrangian constructed by this marginal operator from a pure 2D massless scalar field theory. On the other hand, we use another dimensionful coupling constant $\lambda$, as the parameter of the well-known irrelevant $T \bar{T}$-deformation \cite{Zamolodchikov:2004ce,Smirnov:2016lqw} to construct the GMS theory.
	\subsection{Marginal  $T \bar{T}$-like deformation in ModMax theory}\label{1.1}
The ModMax theory is described by the following Lagrangian density \cite{Bandos:2020jsw}
	\begin{eqnarray}\label{LMM1}
	{\cal L}_{MM}=\cosh(\gamma) \,\mathcal{S}+\sinh(\gamma) \sqrt{ \mathcal{S}^2+\mathcal{P}^2},
	\end{eqnarray}
with two Lorentz invariant scalars $ \mathcal{S}=-\frac{1}{4}F_{\mu\nu}F^{\mu\nu}$ and $\mathcal{P}=-\frac{1}{4}F_{\mu\nu}\tilde F^{\mu\nu}$, where $\tilde{F}^{\mu \nu}= \frac{1}{2} \varepsilon^{\mu \nu\alpha \beta} F_{\alpha \beta}$ is  the Hodge dual of $F^{\mu \nu}$. It was proposed as a unique conformal non-linear theory, which is a marginal modification of the Maxwell electrodynamics \cite{Bandos:2020jsw}. Actually, we have shown in \cite{Babaei-Aghbolagh:2022MoxMax} that there exists a marginal $T \bar{T}$-like operator in the form of \reef{Ogama} that reproduces the ModMax theory from Maxwell theory (see e.g. \cite{Ferko:2022iru}). In the context of  $T \bar{T}$-deformation, we showed that the ModMax as well as the generalized Born-Infeld (GBI) \footnote{ The standard form of the GBI Lagrangian density in $D=4$ is
		\begin{equation}\label{GBI1}
		{\cal L}_{ \gamma BI} = \frac{1}{\lambda} \Bigg[ 1 -  \sqrt{1 -  \lambda \Bigl( 2
			\big( \cosh(\gamma) \mathcal{S}+\sinh(\gamma) \sqrt{ \mathcal{S}^2+\mathcal{P}^2} \big) +\lambda \mathcal{P}^2 \Bigr)} \Bigg],\nn
		\end{equation}
		where in the weak coupling limit $\lambda\to 0$, it reduces to the ModMax theory denoted by \reef{LMM1}.} Lagrangian densities fulfill the flow equations
	\begin{equation}
	\frac{\partial {\cal L}_{MM}}{\partial \gamma}=\frac12\, O_{T^2}^{\gamma},\,\,\,\,\,\,\,\,\,\,\,\,\frac{\partial {\cal L}_{\gamma BI }}{\partial \gamma}=\frac12\, O_{T^2}^{\gamma}\,.
	\end{equation}
	
	Since $\gamma$ is a dimensionless constant parameter, $O_{T^2}^{\gamma}$ is a marginal operator. Therefore, one can perturbatively obtain the expansion of the ModMax theory by applying the marginal $T \bar{T}$-like deformation on the Maxwell theory to desired order. For instance, up to the first order we have
	\begin{equation}
	\label{TTD}
	{\cal L}_{MM} = {\cal L}_{Max}+\frac12 \int O_{T^2}^{\gamma} \,d\gamma.
	\end{equation}
In addition, it was shown \cite{Babaei-Aghbolagh:2022MoxMax} that the electrodynamic GBI theory can be constructed by applying the marginal $T \bar{T}$-like deformation on the BI Lagrangian density 
\begin{equation}
{\cal L}_{BI }  =\frac{1}{\lambda} \bigg[ 1 -  \sqrt{1 -  \lambda \bigl( 2 \mathcal{S}+\lambda \mathcal{P}^2 \bigr)} \bigg].
\end{equation}
	\subsection{2D marginal $T \bar{T}$-like deformation}
	In the general construction of the $T \bar{T}$-deformation in the perturbative approach \cite{Babaei-Aghbolagh:2020kjg}, one should first compute the energy-momentum tensor of the free theory and then find the marginal $T \bar{T}$-like operator $O_{T^2}^{\gamma}={O}^\gamma_{0}$ at the order of $\gamma^0$. This marginal  $T \bar{T}$ operator yields the deformed Lagrangian up to the first order of $\gamma$ as follows
	\begin{equation}
	\label{L1}
	{\cal L}'_{\gamma}={\cal L}_{free}+{\cal L}_1={\cal L}_{free}+ a \int O^\gamma_0 \,d\gamma\,,
	\end{equation}
	where $a$ is a constant coefficient. Using the Lagrangian ${\cal L}'_{\gamma}$ to find the energy-momentum tensor and then substituting in Eq. \eqref{Ogama} to obtain the marginal operator, one can similarly deform the Lagrangian up to the order $\gamma^2$ as
	\begin{equation}
	\label{L2}
	{\cal L}''_{\gamma}={\cal L}_{free}+{\cal L}_1+{\cal L}_2={\cal L}_{free}+ a \int O^\gamma_0 \,d\gamma+ a \int O^\gamma_1 \,d\gamma.
	\end{equation}
	That is, to this order one should use the operator ${O}^\gamma_{T^2}={O}^\gamma_{0}+O^\gamma_1$ to deform the free theory. By iterating this prescription, we can deform the free theory to higher arbitrary orders of $\gamma$.
	In particular, we define the general form of a marginal operator in 2D spacetime by
	\begin{eqnarray}\label{Ogama2}
	O^{\gamma }_{T^2}\equiv\sqrt{T_{\mu\nu}T^{\mu\nu}- \,b\, {T_{\mu}}^{\mu} {T_{\nu}}^{\nu}},
	\end{eqnarray}
	here the coefficient $b$ is a constant. 
	
	Let us consider a free massless scalar theory described by the following Lagrangian
	\begin{eqnarray}\label{pureboson}
	{\cal L}_{ST}=-\frac12 \partial_{\alpha}\Phi \, \partial^{\alpha} \Phi\,,
	\end{eqnarray}
	where ``ST'' stands for scalar theory.
	The corresponding  energy-momentum tensor is given by
	\begin{equation}
	\label{EMT20}
	T_{\mu\nu}(\gamma^{0})=\partial_{\mu}\Phi\,\partial_{\nu}\Phi-\frac12 g_{\mu\nu} \,\partial_{\alpha}\Phi \, \partial^{\alpha} \Phi\,.
	\end{equation}
	It is obvious that the energy-momentum tensor \eqref{EMT20} is traceless which is consistent with the conformal condition for a non-interacting 2D QFT, therefore the constant $b$ does not appear in the marginal operator $O^{\gamma}_{0}$. Now, from \eqref{EMT20} and \eqref{Ogama2} we have
	\begin{equation}
	O^{\gamma}_{T^2}(\gamma^0) \equiv O^{\gamma}_{0} = \frac{1}{\sqrt{2}} \partial_{\alpha}\Phi \, \partial^{\alpha} \Phi,
	\end{equation}
	therefore, the deformed Lagrangian up to the first order of $\gamma$ from Eq. \reef{L1} is given by
	\begin{equation}
	\label{dL1}
	{\cal L}'_\gamma=-\frac12 \partial_{\alpha}\Phi \, \partial^{\alpha} \Phi+\frac{a}{\sqrt{2}}\gamma\, \partial_{\alpha}\Phi \, \partial^{\alpha} \Phi.
	\end{equation}
	
	We note that one can deform the Lagrangian to the next order of $\gamma$ using the first order deformed Lagrangian \reef{dL1}.
	 Consequently, the energy-momentum tensor of the first order deformed theory is obtained as follows
	\begin{equation}
	\label{EMT21}
	T_{\mu \nu}(\gamma) =(1 -\sqrt{2} \, a\,  \gamma) \partial_{\mu}\Phi\,\partial_{\nu}\Phi + (\frac{a}{\sqrt{2}} \gamma- \frac{1}{2})\, \mathit{g}_{\mu \nu}\, \partial_{\alpha}\Phi \, \partial^{\alpha} \Phi\, =(1 -  \sqrt{2}\, a \,\gamma) \, T_{\mu\nu}(\gamma^{0})\,.
	\end{equation}
	The contribution of \eqref{dL1} to $T\bar{T}$ operator due to ${O}^\gamma_{T^2}={O}^\gamma_{0}+O^\gamma_1$ at the order of $\gamma$ is
	\begin{equation}\label{dL22}
	O^\gamma_{T^2}(\gamma)=\frac{a}{\sqrt{2}}  \,\partial_{\alpha}\Phi \, \partial^{\alpha} \Phi -    a^2 \, \gamma\, \partial_{\alpha}\Phi \, \partial^{\alpha} \Phi\,,
	\end{equation}
	so that by integrating \reef{dL22}, we obtain the marginal second-order deformed Lagrangian as
	\begin{equation}
	\label{dL2}
	{\cal L}''_{\gamma}= - \frac{1}{2}\partial_{\alpha}\Phi \, \partial^{\alpha} \Phi + \frac{a}{\sqrt{2}} \, \gamma \, \partial_{\alpha}\Phi \, \partial^{\alpha} \Phi  -  \frac{a^2}{2}\, \gamma^2\, \partial_{\alpha}\Phi \, \partial^{\alpha} \Phi \,.
	\end{equation}
	Iterating this procedure, for example one can reach to the following marginally deformed Lagrangian of the order $\gamma^5$
		\begin{eqnarray}
	\label{dL5}
	{\cal L}^5_{\gamma}&\!\!\!=\!\!\!& - \frac{1}{2}\partial_{\alpha}\Phi \, \partial^{\alpha} \Phi + \frac{a}{\sqrt{2}} \, \gamma \, \partial_{\alpha}\Phi \, \partial^{\alpha} \Phi  -  \frac{a^2}{2}\, \gamma^2\, \partial_{\alpha}\Phi \, \partial^{\alpha} \Phi \\
	&\!\!\!+\!\!\!& \frac{a^3}{3 \sqrt{2}} \, \gamma^3 \, \partial_{\alpha}\Phi \, \partial^{\alpha} \Phi -  \frac{a^4}{12} \, \gamma^4 \, \partial_{\alpha}\Phi \, \partial^{\alpha} \Phi + \frac{a^5}{30 \sqrt{2}} \, \gamma^5 \, \partial_{\alpha}\Phi \, \partial^{\alpha} \Phi\,,\nonumber
	\end{eqnarray}
	nonetheless, after a straightforward calculation we find that the deformed Lagrangian to all orders is equivalent to the expansion of the bellow Lagrangian
	\begin{eqnarray}\label{dLgmma}
	{\cal L}_{\gamma}=-\frac{1}{2}\big(\cosh(\sqrt{2} \,a \, \gamma \,) - \sinh(\sqrt{2} \,a \, \gamma \,)\big) \partial_{\alpha}\Phi \, \partial^{\alpha} \Phi\,.
	\end{eqnarray}
	
	Now, if we fix the constant $a$ as $a=-\frac{1}{\sqrt{2}}$ and use the relations  $\cosh(-\gamma)=\cosh(\gamma),\,\,\,\sinh(-\gamma)=-\sinh(\gamma)$, the Lagrangian density \eqref{dLgmma} recast to  
	\begin{equation}
	\label{SMMd}
	{\cal L}_{MS}(\gamma)=-\frac{1}{2}\big(\cosh( \gamma ) + \sinh( \gamma )\big) \partial_{\alpha}\Phi \, \partial^{\alpha} \Phi,
	\end{equation}
	which is called the marginal scalar (MS) theory. The energy-momentum tensor of the MS theory \eqref{SMMd} is
		\begin{eqnarray}
	\label{Tgama}
	T_{\mu\nu} =e^{\gamma }\,\partial_{\mu}\Phi\,\partial_{\nu}\Phi -\frac{1}{2}\, e^{\gamma} \mathit{g}_{\mu \nu} \, \partial_{\alpha}\Phi \, \partial^{\alpha} \Phi \,,
	\end{eqnarray}
	thus, the corresponding marginal operator becomes
\begin{equation}
	\label{ogsm}
	O_{T^2}^{\gamma}= \frac{1}{\sqrt{2}} e^\gamma \partial_{\alpha}\Phi \, \partial^{\alpha} \Phi.
	\end{equation}
	 It can be easily checked that the operator \eqref{ogsm} and the MS Lagrangian \eqref{SMMd} satisfy the following flow equation
	\begin{equation}
	\label{TTyD}
	\frac{\partial {\cal L}_{MS}(\gamma)}{\partial \gamma}=-\frac{1}{ \sqrt{2}}\, O_{T^2}^{\gamma},
	\end{equation}
	which is the main implication in the $T \bar{T}$-like deformation of a QFT.
	\subsection{Irrelevant $T \bar{T}$-deformation of 2D MS theory}
	In the following, we study the deformation of the MS theory \reef{SMMd} under an irrelevant $T \bar{T}$ operator. According to the perturbative approach in \cite{Babaei-Aghbolagh:2020kjg} relative to the standard irrelevant $T \bar{T}$-deformation with constant parameter $\lambda$ in Refs. \cite{Zamolodchikov:2004ce,Smirnov:2016lqw}, we deform the MS Lagrangian up to the first order of $\lambda$ with
	\begin{equation}
	\label{irreleventTTD}
	{\cal L'}_\lambda ={\cal L}_{ MS}(\, \gamma \,) +\frac{1}{8} \int O_{0}^{\lambda} \,d\lambda\,,
	\end{equation}
	where $O_{0}^{\lambda}$ is the zeroth order of the irrelevant  $T \bar{T}$ operator
	\begin{eqnarray}
	\label{olamda}
	O_{T^2}^{\lambda} = T_{\mu\nu}T^{\mu\nu}- {T_{\mu}}^{\mu} {T_{\nu}}^{\nu}\,,
	\end{eqnarray}
	and $T_{\mu\nu}$ is the energy-momentum tensor in \reef{Tgama}. Thus, up to the first order of $\lambda$ the deformed MS Lagrangian becomes
	\begin{equation}
	\label{irrelev1}
	{\cal L'}(\gamma,\lambda) =- \frac{1}{2 } e^{\gamma} \, \partial_{\alpha}\Phi \, \partial^{\alpha} \Phi + \frac{1}{16}\,  e^{2\gamma}\,\lambda  \, (\partial_{\alpha}\Phi \, \partial^{\alpha} \Phi)^2.
	\end{equation}
	By iterating this method, one can derive the irrelevant $T \bar{T}$-deformed MS Lagrangian to higher orders of $\lambda$
	that at each order, is exactly equal to the expansion of the generalized marginal scalar (GMS) Lagrangian given by
	\begin{eqnarray}
	\label{irrel}
	{\cal L}_{GMS}(\gamma,\lambda) = \frac{2}{\lambda}  \left(1 -  \sqrt{1 + \tfrac{1}{2}\, \lambda \, \big( \cosh(\gamma) +\sinh(\gamma)  \big) \, \partial_{\alpha}\Phi \, \partial^{\alpha} \Phi}\,\right)=\frac{2}{\lambda}  \left(1 -  \sqrt{1 + \tfrac{1}{2}\, e^{\gamma}\, \lambda \, \partial_{\alpha}\Phi \, \partial^{\alpha} \Phi}\,\right).
	\end{eqnarray}
	In fact, the GMS Lagrangian \reef{irrel} is a 2D general Lagrangian with two coupling constants $\gamma$ and $\lambda$, such that in the limit $\gamma \to 0$, it reduces to the scalar BI-like Lagrangian as
	\begin{equation}
	\label{BI2}
		{\cal L}_{SBI}(\lambda)=\frac{2}{\lambda} \left(1-\sqrt{1+\tfrac12 \lambda\,\partial_{\alpha}\Phi \, \partial^{\alpha} \Phi}\right),
	\end{equation}
and in the limit $\lambda \to 0$, reduces to the MS theory given in Eq. \reef{SMMd}.
	
In summary, the scalar field theories in $D=2$ can be deformed by two different types of $T\bar{T}$-deformation; the irrelevant and the marginal deformations. We followed these two $T\bar{T}$-deformations by a perturbative approach and found the GMS Lagrangian \reef{irrel}. In other words, for a deformed scalar theory, there are two types of $T \bar{T}$-like flow in $D=2$.
	
On the other hand, the marginal $T\bar{T}$ operator introduced in Eq. \reef{Ogama2} includes an undetermined constant coefficient $b$ that should be fixed appropriately. To this end, one needs to solve the following differential equation

		\begin{equation}\label{blamda}
		\frac{\partial  {\cal L}_{GMS}}{\partial \gamma}=-\frac{1}{\sqrt{2}}\, \sqrt{T_{\mu\nu}T^{\mu\nu}-\,b\, {T_{\mu}}^{\mu} {T_{\nu}}^{\nu}} \,.
	\end{equation}
	The left hand side of this equation is easily computed from the GMS Lagrangian \reef{irrel}, so we obtain
	\begin{equation}\label{Tga}
\frac{\partial  {\cal L}_{GMS}}{\partial \gamma}=- \frac{e^\gamma \, \partial_{\alpha}\Phi \, \partial^{\alpha} \Phi}{2 \sqrt{1 + \frac{1}{2}\,\lambda\, e^\gamma  \, \partial_{\alpha}\Phi \, \partial^{\alpha} \Phi}} \,,
	\end{equation}
	while the energy-momentum tensor on the right hand side for the GMS theory is obtained as follows
	\begin{eqnarray}
	\label{Tgamalamda}
	T_{\mu\nu} =\frac{e^\gamma \lambda \,\partial_{\mu}\Phi\,\partial_{\nu}\Phi - 2 \left(1+  \tfrac{1}{2} e^\gamma \lambda \, \partial_{\alpha}\Phi \, \partial^{\alpha} \Phi - \sqrt{1 + \tfrac{1}{2} e^\gamma \lambda \, \partial_{\alpha}\Phi \, \partial^{\alpha} \Phi}\,\right) \mathit{g}_{\mu \nu}}{\lambda\, \sqrt{1 + \tfrac{1}{2} e^\gamma \lambda \, \partial_{\alpha}\Phi \, \partial^{\alpha} \Phi}}.
	\end{eqnarray}
	By substituting the relations \reef{Tga} and \reef{Tgamalamda} in Eq. \reef{blamda}, we obtain $b=\frac{1}{2}$.	Therefore, there are two types of $T \bar{T}$-like flows for the GMS theory;  the irrelevant (with respect to $\lambda$) and the marginal (with respect to $\gamma$). These two $T \bar{T}$-like flows are described by
	\begin{equation}\label{lM}
	\frac{\partial {\cal L}_{GMS}}{\partial \lambda}=\frac18 \left( T_{\mu\nu}T^{\mu\nu}- {T_{\mu}}^{\mu} {T_{\nu}}^{\nu}\right)\,,\qquad \frac{\partial  {\cal L}_{GMS}}{\partial \gamma}=-\frac{1}{\sqrt{2}}\, \sqrt{T_{\mu\nu}T^{\mu\nu}-\tfrac{1}{2} {T_{\mu}}^{\mu} {T_{\nu}}^{\nu}} \,.
	\end{equation}	
	\section{Scalar ModMax theory from marginal  $T \bar{T}$ -deformation} \label{3}
	
	In previous section, we introduced a marginal operator that deforms a pure 2D scalar field theory into the MS theory. Though the marginal operator has a non-trivial form, the  structure of the resulting MS theory \eqref{SMMd} is the same as initial theory with an additional scaling factor $e^\gamma$. In this section, we will investigate the role of the marginal operator \eqref{Ogama2} in deforming a theory of non-interacting multi-scalar fields. Using the perturbative method, we show that the emergent theory is a non-trivial action of massless scalar fields which has the form of ModMax theory but in $D=2$. We name this modified action as the scalar ModMax (SMM) theory.
	
	 Let us consider a theory with $N \geq  1$ scalar fields $\Phi^i$, where $i=1,2, \dots,N$. The Lagrangian of this multi-scalar theory is described by
	 \begin{equation}\label{MS}
	 {\cal L}_{MST}=-\frac12 \partial_{\alpha}\Phi^i \, \partial^{\alpha} \Phi_i\,,
	 \end{equation}
	 One can apply the marginal $T \bar{T}$ deformation by operator \eqref{Ogama2} to the Lagrangian \eqref{MS} as the free theory. From perturbative approach, up to the first order of $\gamma$ we obtain
	  \begin{equation}
	  \label{LMS}
	  {\cal L}'(\gamma)={\cal L}_{MST}+{\cal L}_1={\cal L}_{MST}-\frac12 \int O^\gamma_0 \,d\gamma,
	  \end{equation}
	  where $O^\gamma_0=-\frac{1}{\sqrt{2}}\, \sqrt{T_{\mu\nu}T^{\mu\nu}-\frac{1}{2} {T_{\mu}}^{\mu} {T_{\nu}}^{\nu}} \,$ is the marginal operator at the order of $\gamma^0$ and $T_{\mu\nu}$ is computed from the Lagrangian \reef{MS} as follows
	   \begin{equation}\label{TMMS}
	  T^{MST}_{\mu \nu}=G_{ij} \partial_{\mu}\Phi^{i} \partial_{\nu}\Phi^{j}- \frac{1}{2} g_{\mu \nu}\, \partial_{\alpha}\Phi_{i} \partial^{\alpha}\Phi^{i}\,.
	  \end{equation}
	  Here, $G_{ij}$ is a symmetric tensor with indices for the moduli space of scalar fields.  Therefore, the deformed Lagrangian density \reef{LMS} is given by
	 \begin{equation}\label{Lga}
      {\cal L}'(\gamma)=	-\frac{1}{2} \partial_{\alpha}\Phi_{i} \partial^{\alpha}\Phi^{i} - \frac{1}{2}  \,\gamma \, \sqrt{- (\partial_{\alpha}\Phi_{i} \partial^{\alpha}\Phi^{i})^2 + 2 (G_{ik} G_{jl} \partial_{\alpha}\Phi^{j} \partial^{\alpha}\Phi^{i} \partial_{\beta}\Phi^{l} \partial^{\beta}\Phi^{k})}.
	 \end{equation}
	 For simplicity, we can define the following two variables
	 \begin{equation}
	 \label{P1P21}
	 P_1=\partial_{\alpha}\Phi_{i} \partial^{\alpha}\Phi^{i},\quad P_2=G_{ik} G_{jl} \partial_{\alpha}\Phi^{j} \partial^{\alpha}\Phi^{i} \partial_{\beta}\Phi^{l} \partial^{\beta}\Phi^{k}\,,
	 \end{equation}
	then the first-order marginally deformed Lagrangian rewritten as
	\begin{equation}\label{Lgap1}
	{\cal L}'(\gamma)=	-\frac{1}{2} P_1 - \frac{1}{2} \,\gamma \, \sqrt{- P_1^2 + 2\, P_2} \,.
	\end{equation}
	
	For the next order, we first calculate the energy-momentum tensor corresponding to ${\cal L}'(\gamma)$, i.e.,
	\begin{eqnarray}\label{Tga2}
	T'_{\mu \nu}(\gamma)= \partial_{\mu}\Phi^{i} \partial_{\nu}\Phi_{i}- \frac{1}{2} g_{\mu \nu} P_1  -  \frac{1}{2} \,\gamma \, g_{\mu \nu} \sqrt{- P_1^2 + 2\, P_2}-\,\gamma \,  \frac{ P_1 \,\partial_{\mu}\Phi^{i} \partial_{\nu}\Phi_{i}}{\sqrt{- P_1^2 + 2\, P_2}}
+\,\gamma \, \frac{4 \, \partial_{\alpha}\Phi_{j} \partial^{\alpha}\Phi_{i} \partial_{\mu}\Phi^{i} \partial_{\nu}\Phi^{j}}{2 \sqrt{- P_1^2 + 2\, P_2}}\,.
	\end{eqnarray}
	Then, by substituting the above energy-momentum tensor into the marginal $T \bar{T}$ operator so that ${O}^\gamma_{T^2}={O}^\gamma_{0}+O^\gamma_1$, one can find the deformed lagrangian up to the order $\gamma^2$ as
	\begin{eqnarray}\label{L2gal}
	{\cal L}''(\gamma)=-\frac{1}{2} P_1 - \frac{1}{2} \,\gamma \, \sqrt{- P_1^2 + 2\, P_2} - \frac{1}{4} \,\gamma^2 \,\frac{ P_1^3 - 4 P_1 P_2 + 4 \,\,\partial_{\alpha}\Phi^{j} \partial^{\alpha}\Phi^{i} \partial_{\beta}\Phi^{k} \partial^{\beta}\Phi_{i} \partial_{\gamma}\Phi_{k} \partial^{\gamma}\Phi_{j}}{- P_1^2+ 2 P_2}\,.
	\end{eqnarray}
	We find that the Lorentz invariant variables $P_1$ and $P_2$ satisfy the following identity \footnote{We have already found similar trace identities in 4D gauge theories in the Appendix of Ref. \cite{BabaeiVelni:2016qea}(also see \cite{Ferko:2206jsw}) }
\begin{eqnarray}\label{iden}
\partial_{\alpha}\Phi^{j} \partial^{\alpha}\Phi^{i} \partial_{\beta}\Phi^{k} \partial^{\beta}\Phi_{i} \partial_{\gamma}\Phi_{k} \partial^{\gamma}\Phi_{j}=- \frac{1}{2}\, P_1^3 + \frac{3}{2}\, P_1\, P_2\,,
	\end{eqnarray}	
	which shows that the deformed Lagrangian depends only on these two trace structures. By replacing the identity \reef{iden} in Eq. \reef{L2gal}, we obtain
		\begin{eqnarray}\label{L22}
	{\cal L}''(\gamma)=-\frac{1}{2} P_1 - \frac{1}{2} \,\gamma \, \sqrt{- P_1^2 + 2\, P_2} - \frac{1}{4} \,\gamma^2 \,P_1\,.
	\end{eqnarray}
 Again one can proceed this method to deform the Lagrangian \eqref{LMS} to higher arbitrary orders of the deformation constant parameter $\gamma$.

 It is shown that the final Lagrangian at all orders of $\gamma$ can be expressed as the following closed form
\begin{equation}
	\label{SMMdw}
	{\cal L}_{SMM}(\gamma)=-\frac{1}{2}\left(\cosh( \gamma )\, P_1+ \sinh( \gamma ) \sqrt{- P_1^2 + 2\, P_2}\,\right) \,,
	\end{equation}
which is the SMM theory in $D=2$. It has been discussed in Ref. \cite{Conti:2022egv} that one can obtain this 2D ModMax-like scalars field theory from dimensional reduction of a 4D ModMax theory. In order to study the flow of the SMM theory, we first should compute the energy-momentum tensor corresponding to SMM Lagrangian \reef{SMMdw}. Then, we obtain
	\begin{eqnarray}
	T_{\mu\nu}&\!\!\!=\!\!\!&- \frac{1}{2}  g_{\mu \nu} \bigg(\cosh(\gamma)\, P_1+ \sinh(\gamma)\,\sqrt{- P_1^2 + 2\, P_2}\bigg) \\
	& \!\!\!+\!\!\!& \bigg( \cosh(\gamma)  -  \frac{\sinh(\gamma) \,P_1}{\sqrt{- P_1^2 + 2\, P_2}}\bigg)\partial_{\mu}\Phi^{i} \partial_{\nu}\Phi_{i}+\frac{2 \sinh(\gamma)\, \partial_{\alpha}\Phi_{j} \partial^{\alpha}\Phi_{i} \partial_{\mu}\Phi^{i} \partial_{\nu}\Phi^{j}}{\sqrt{- P_1^2 + 2\, P_2}},\nonumber
	 \end{eqnarray}
	 which is traceless. The SMM theory also satisfies the flow equation $\frac{\partial  {\cal L}^{SMM}}{\partial \gamma}=-\frac{1}{\sqrt{2}}\, \sqrt{T_{\mu\nu}T^{\mu\nu}-\frac{1}{2} {T_{\mu}}^{\mu} {T_{\nu}}^{\nu}} \,$.
	
It is of interest to also investigate the flow of the SMM theory under the effect of applying an irrelevant $T \bar{T}$-deformation. In this respect, one can deform the SMM Lagrangian order by order to reproduce a general Lagrangian of this theory similar to that in Eq. \reef{irrel} obtained for GMS model. For example, up to the order of $\lambda^3$, the deformed SMM theory is given by
\begin{eqnarray}
{\cal L}^{\lambda^3}&\!\!\!=\!\!\!& {\cal L}_{SMM} +\frac{1}{8} \lambda   (2 {\cal L}_{SMM}^2 -  P_1^2 + P_2) +\frac{1}{16} \lambda^2 {\cal L}_{SMM}  (2 {\cal L}_{SMM}^2 -  P_1^2 + P_2) \\
&\!\!\!+\!\!\!&  \frac{1}{256} \lambda^3 (2 {\cal L}_{SMM}^2 -  P_1^2 + P_2) (10 {\cal L}_{SMM}^2 -  P_1^2 + P_2).\nonumber
\end{eqnarray}
This expression is exactly equal to the expansion of the GSMM theory
	\begin{eqnarray}\label{GSMMdw}
      {\cal L}_{GSMM} =\frac{2}{ \lambda} \bigg(1 - \sqrt{1- \lambda {\cal L}_{SMM}+\frac{1}{8} \lambda^2 (P_1^2 -  P_2)}\bigg),
    \end{eqnarray}
   up to corresponding order of $\lambda$. Note that in the limit $\gamma \to 0$, the GSMM Lagrangian \eqref{GSMMdw} reduces to the following theory
    \begin{eqnarray}\label{Msbi}
    {\cal L}_{MSBI} =\frac{2}{ \lambda} \bigg(1 - \sqrt{1+\tfrac{1}{2} \lambda P_1+\frac{1}{8} \lambda^2 (P_1^2 -  P_2)}\bigg),
    \end{eqnarray}
   where ``MSBI'' stands for multi-scalar BI-like theory. In other words, the GSMM theory is generated from deformation of the multi-scalar BI-like theory in Eq. \reef{Msbi} by using the marginal $T \bar{T}$-like operator \eqref{Ogama2}. In breif, one can examinate that the GSMM Lagrangian density \eqref{GSMMdw} fulfill both the irrelevant and the marginal flow equations as follows
	\begin{equation}\label{lM2}
	\frac{\partial {\cal L}_{GSMM}}{\partial \lambda}=\frac18\, \bigg( T_{\mu\nu}T^{\mu\nu}- {T_{\mu}}^{\mu} {T_{\nu}}^{\nu}\bigg)\,,\,\,\,\,\,\,\,\,\,\,\,\,\frac{\partial  {\cal L}_{GSMM}}{\partial \gamma}=-\frac{1}{\sqrt{2}}\, \sqrt{T_{\mu\nu}T^{\mu\nu}-\frac{1}{2} {T_{\mu}}^{\mu} {T_{\nu}}^{\nu}} \,.
	\end{equation}
\section{Conclusion and outlook}\label{4}
The fact which how we can add the interaction terms to a free QFT has always been of interest. Actually these terms are added to the free theory by a coupling constant so that in the weak coupling limit, the free theory is revived. Irrelevant $T \bar{T}$-deformation is one of the approaches that deforms a free theory into an interacting version. This deformed theory is solvable in $D=2$ but does not respect conformal symmetry. In this paper, we proposed another type of $T \bar{T}$-deformation which respects conformal symmetry. We have shown that there is a marginal operator that reproduces a deformed theory of 2D scalar fields with conformal symmetry. In particular, by applying the marginal $T \bar{T}$ operator \eqref{Ogama2} on a pure scalar and multi-scalar theories described by Eqs. \reef{pureboson} and \reef{MS}, we reproduced the 2D MS and SMM theories respectively by \reef{SMMd} and \reef{SMMdw}. 

On the other hand, we have also considered the effect of the irrelevant $T \bar{T}$-deformation on the marginally deformed actions. We obtained the generalized BI-like Lagrangian densities for these theories. It was proved that these two types of $T \bar{T}$-deformation operators are commute with each other. Also it seems that these two $T \bar{T}$-like flows could be applied to any QFTs. In order to summarize our results here, we have depicted diagrammatically the effects of two irrelevant and marginal operators ${O}_{T^2}^{\lambda}$ and $O_{T^2}^{\gamma}$ on SM and MST theories in Fig. \ref{fig1}.

\begin{center}
	\begin{tikzcd}
	{\cal L}_{ST} \arrow[r, blue, "{O}_{T^2}^{\lambda}" blue] \arrow[d,red,"O_{T^2}^{\gamma}" red]
	&|[blue]| {\cal L}_{SBI} \arrow[d,red, "O_{T^2}^{\gamma}" red] \\
	|[red]|{\cal L}_{MS} \arrow[r, blue, "{O}_{T^2}^{\lambda}" blue]
	&|[red!50!blue]|  {\cal L}_{GMS}
	\end{tikzcd}
	\hspace{2cm}
	\begin{tikzcd}
		{\cal L}_{MST} \arrow[r, blue, "{O}_{T^2}^{\lambda}" blue] \arrow[d,red,"O_{T^2}^{\gamma}" red]
		&|[blue]| {\cal L}_{MSBI} \arrow[d,red, "O_{T^2}^{\gamma}" red] \\
		|[red]|{\cal L}_{SMM} \arrow[r, blue, "{O}_{T^2}^{\lambda}" blue]
		&|[red!50!blue]|  {\cal L}_{GSMM}
		\end{tikzcd}
	\captionof{figure}{Deformations of the free scalar and multi-scalar theories under ${O}_{T^2}^{\lambda}$ and $O_{T^2}^{\gamma}$.}\label{fig1}
\end{center}

In order to better understand the implication of the marginal deformation, we need to study more theories under this deformation. 2D fermionic and non-relativistic theories are topics of interest for further studies. Also, the structure of the marginal $T \bar{T}$-deformation in higher dimensions would be topics of our interest.

   \vspace{0.6cm}
   \noindent
   {\bf Note Added:}
 While we were in the final stage of this work, the paper \cite{Ferko:2206jsw} appeared in the arXiv where the modification of 2D scalar field theory has also been studied as a root flow equation which has some overlap with our results on the scalar ModMax theory.	
	
	\section*{Acknowledgment}
	The authors would also like to thank M. R. Garousi, M.M. Sheikh-jabbari, A. Ghodsi, and H. R. Afshar for valuable discussions. We are also grateful to Dmitri Sorokin for fruitful discussion and helpful comments on the generalized ModMax theories. HBA is specially appreciate Song He for his useful comments at the early stages of the work.
	
	\if{}
	\bibliographystyle{abe}
	\bibliography{references}{}
	\fi
	
	\providecommand{\href}[2]{#2}\begingroup\raggedright\endgroup
\end{document}